# Angular momentum of the electromagnetic field: the plane wave paradox resolved

**A. M. Stewart**

Research School of Physical Sciences and Engineering,
The Australian National University,
Canberra, ACT 0200, Australia.

**Abstract.** The angular momentum of a classical electromagnetic plane wave of arbitrary extent is predicted to be, on theoretical grounds, exactly zero. However, finite sections of circularly polarized plane waves are found experimentally to carry angular momentum and it is known that the contribution to the angular momentum arises from the edges of the beam. A mathematical model is described that gives a quantitative account of this effect and resolves the paradox.

## 1. Introduction

The angular momentum $\boldsymbol{J}(t)$ of the classical electromagnetic field is given in terms of the electric $\boldsymbol{E}(\boldsymbol{r},t)$ and magnetic $\boldsymbol{B}(\mathbf{r},t)$ Maxwell fields by [1]

$$\boldsymbol{J}(t) = \frac{1}{4\pi c}\int d^3r\, \boldsymbol{r} \times [\boldsymbol{E}(\boldsymbol{r},t) \times \boldsymbol{B}(\boldsymbol{r},t)] \tag{1}$$

(Gaussian units, bold font denotes a three-vector). For a plane wave propagating in the $z$ direction the electric and magnetic fields lie in the $x$-$y$ plane. Accordingly, the Poynting vector $\boldsymbol{S} = \boldsymbol{E}\times\boldsymbol{B}$, the vector cross product of $\boldsymbol{E}$ and $\boldsymbol{B}$, lies also in the $z$ direction. Because of this, the component of the angular momentum in the direction of propagation must be zero, due to the vector cross product of $\boldsymbol{r}$ with $\boldsymbol{E}\times\boldsymbol{B}$ in (1). For paraxial waves, which comprise a beam of limited radius such as that produced by a laser, this is clearly not the case and the angular momentum properties of these [2-5] have been analyzed in detail.

However, it has been known ever since the experiments of Beth [6] that a circularly polarized plane wave of light does carry angular momentum. This apparent paradox has been the subject of discussion for a long time [7-9] and even recently [10-12]. The resolution of the paradox is that any obstacle that absorbs the beam changes the electromagnetic field at the edges of the obstacle so as to give the fields in that region components in the direction of propagation. These give rise to a Poynting vector that is azimuthal with respect to the direction of propagation and thus to an angular momentum vector parallel to it. Some discussion of the matter has been given by Simmons and Guttman [13]. It seems that although a circularly polarized plane wave of arbitrary extent may not carry angular momentum in an *actual* form it does carry it in a *potential* form in that any constriction of the wave, such as by an aperture, will give rise to an angular momentum.

The decomposition of the angular momentum of general (i.e. non-paraxial) electromagnetic waves has been approached by some authors. Ohanian [14] treated the angular momentum by expressing the magnetic field as the curl of the vector potential and decomposing the latter, but his method has the disadvantage of not being gauge invariant. Barnett and Allen [15] assumed an azimuthal phase dependence of one particular form. Barnett [16] applied the concept of angular momentum flux arising from the conservation laws of classical field theory to monochromatic waves of specified azimuthal phase dependence but did not treat the boundary conditions necessary to resolve the plane wave issue.

This matter is of interest for the teaching and learning of electromagnetism. A plane wave is the simplest example of an electromagnetic field with non-trivial properties and widespread practical applications. A student can readily deduce the linear momentum density of a plane wave from its Poynting vector. However, corresponding arguments for the angular momentum are open to question, as detailed above, and do not appear to be discussed adequately in current texts on electromagnetism. We provide a straightforward resolution of the issue by showing that the angular momentum of a given volume of the free electromagnetic field may be decomposed into three terms: a volume integral of a spin-like term (the first term of equation 15), the volume integral of a orbital-like term (equation 17) and a surface integral (the second term of equation 15). If the field is localized in a finite volume then the volume of integration may be extended to infinity, where the fields vanish, and the surface integral will be zero. However, for a plane wave, the fields will always be significant at the surface of the volume considered and the surface integral must be taken account of. When this is done the paradox is resolved.

It is the purpose of this paper to provide an exact and manifestly gauge invariant treatment of the angular momentum of general electromagnetic waves with particular emphasis on the effect of boundaries on the plane wave problem. We do this by extending our previous decomposition of the angular momentum of the electromagnetic field by means of the Helmholtz theorem [17-20] to take account of the surface integrals that inevitably arise when dealing with a wave that is of arbitrary extent such as a plane wave. In section 2 of this paper the theory is established and in section 3 the results are applied to plane waves of arbitrary polarization. Section 4 concludes the paper.

## 2. Theory

It has been shown elsewhere [17] that by expressing the electric field vector as the sum of its longitudinal and transverse components by means of the Helmholtz theorem

$$\boldsymbol{E}(\boldsymbol{r},t) = \quad \times \int \mathrm{d}V' \frac{' \times \boldsymbol{E}(\boldsymbol{r}',t)}{4\pi|\boldsymbol{r}-\boldsymbol{r}'|} - \quad \int \mathrm{d}V' \frac{' . \boldsymbol{E}(\boldsymbol{r}',t)}{4\pi|\boldsymbol{r}-\boldsymbol{r}'|} \tag{2}$$

where ' is the gradient operator with respect to ***r***'. The validity of using the Helmholtz decomposition for 3-vector fields that depend on time has been justified elsewhere [19, 21]. Using Maxwell's equations we get

$$\boldsymbol{E}(\boldsymbol{r},t) = \quad \times \boldsymbol{F} - \quad \int \mathrm{d}^3 y \frac{\rho(\boldsymbol{y},t)}{|\boldsymbol{r}-\boldsymbol{y}|} \tag{3}$$

with

$$\boldsymbol{F}(\boldsymbol{r},t) = -\int \mathrm{d}^3 y \frac{\partial \boldsymbol{B}(\boldsymbol{y},t)/\partial t}{4\pi c\,|\boldsymbol{r}-\boldsymbol{y}|} \tag{4}$$

where ρ(**y**,*t*) is the electric charge density. The two terms of (3) are substituted into (1) to give two contributions to the angular momentum. The second term of (3), that involving the charge density, is relevant to fields in the vicinity of charge density. Its properties have been examined elsewhere [17] and will not discussed here. The first term of (3), which relates to the properties of free fields and whose divergence is zero (the transverse component of ***E***), gives a contribution of

$$\boldsymbol{J}_{\mathrm{f}} = \frac{1}{4\pi c}\int \mathrm{d}^3 r\, \boldsymbol{r} \times [(\nabla \times \boldsymbol{F}) \times \boldsymbol{B}] \tag{5}$$

to the angular momentum. Although in the next section of the paper we concentrate on plane waves, the term *free fields* used in the present section refers to wave fields of arbitrary spatial and time dependence.

By making use of the standard vector identity

$$(\nabla \times \boldsymbol{F}) \times \boldsymbol{B} = (\boldsymbol{B}.\nabla)\boldsymbol{F} - \sum_{r=1}^{3} B^r \nabla F^r \tag{6}$$

(5) is divided into two parts. We first consider the first one $\boldsymbol{J}_{\mathrm{fs}}$ coming from the first term of (6). By using the identity

$$(\boldsymbol{B}.\nabla)(\boldsymbol{r} \times \boldsymbol{F}) = \boldsymbol{B} \times \boldsymbol{F} + \boldsymbol{r} \times (\boldsymbol{B}.\nabla)\boldsymbol{F} \quad , \tag{7}$$

which may be verified by expanding in Cartesian components, we get

$$\boldsymbol{J}_{\mathrm{fs}} = \frac{1}{4\pi c}\int \mathrm{d}^3 r\, \boldsymbol{F} \times \boldsymbol{B} + \frac{1}{4\pi c}\int \mathrm{d}^3 r\, (\boldsymbol{B}.\nabla)\boldsymbol{G} \tag{8}$$

where ***G*** = ***r***x***F***.

To clarify the nature of the second term of (8) we introduce the tensor $\overleftrightarrow{\boldsymbol{T}} = \boldsymbol{BG}$, written in dyadic form,

$$\overleftrightarrow{\boldsymbol{T}} = \boldsymbol{BG} = \sum_{r,i=1}^{3} \hat{\varepsilon}_r B^r G^i \hat{\varepsilon}_i \quad , \qquad (9)$$

where $\hat{\varepsilon}_r$ is the unit vector in the $r$ direction. We consider the divergence of $\overleftrightarrow{\boldsymbol{T}}$

$$\ .\overleftrightarrow{\boldsymbol{T}} = \sum_{ri} \hat{\varepsilon}_r . \hat{\varepsilon}_r \frac{\partial}{\partial x^r} (B^r G^i) \hat{\varepsilon}_i \quad . \qquad (10)$$

Taking the derivative explicitly and remembering that $\ .\boldsymbol{B} = 0$ we get

$$\ .\overleftrightarrow{\boldsymbol{T}} = (\boldsymbol{B}.\ \ )\boldsymbol{G} \qquad (11)$$

and find that the second volume integral in (8) is that of the divergence of the tensor $\overleftrightarrow{\boldsymbol{T}}$

$$\frac{1}{4\pi c} \int \mathrm{d}^3 r \ \ .\overleftrightarrow{\boldsymbol{T}} \quad . \qquad (12)$$

Using Gauss's law applied to a tensor

$$\int_V \mathrm{d}^3 r \ \ .\overleftrightarrow{\boldsymbol{T}} = \int_S \mathrm{d}\boldsymbol{s}.\overleftrightarrow{\boldsymbol{T}} \quad , \qquad (13)$$

where d$\boldsymbol{s}$ is a directed infinitesimal area of the surface that bounds the volume, we can express (8) as the sum of a volume integral and a surface integral

$$\boldsymbol{J}_{\mathrm{fs}} = \frac{1}{4\pi c} \int_V \mathrm{d}^3 r \boldsymbol{F} \times \boldsymbol{B} + \frac{1}{4\pi c} \int_S (\mathrm{d}\boldsymbol{s}.\boldsymbol{B}) \boldsymbol{G} \quad . \qquad (14)$$

Written out explicitly in terms of the fields this is

$$\boldsymbol{J}_{\mathrm{fs}} = \frac{1}{(4\pi c)^2}\int_V \mathrm{d}^3x \int_V \mathrm{d}^3y \, \frac{\boldsymbol{B}(\boldsymbol{x},t)}{|\boldsymbol{x}-\boldsymbol{y}|} \times \frac{\partial \boldsymbol{B}(\boldsymbol{y},t)}{\partial t} - \frac{1}{(4\pi c)^2}\int_S \mathrm{d}^2\boldsymbol{x}.\boldsymbol{B}(\boldsymbol{x},t) \int_V \mathrm{d}^3y \, \boldsymbol{x} \times \frac{\partial \boldsymbol{B}(\boldsymbol{y},t)}{\partial t} \frac{1}{|\boldsymbol{x}-\boldsymbol{y}|} \quad (15)$$

where $\mathrm{d}^2\boldsymbol{x}$ is the directed surface area element of $\boldsymbol{x}$. The first term $\boldsymbol{J}_{\mathrm{fs}}$ (15) has the nature of a spin angular momentum because the coordinate vector does not appear explicitly in the numerator of the integrals. For fields that vanish sufficiently fast at infinity the second term of (15) can be ignored if the volume of integration is extended to infinity, but for fields that do not, such as a plane wave of arbitrary extent, it clearly cannot.

The second term of (6), when substituted into (1) cannot, by repeated partial integrations, be cast into a form that does not depend linearly on the coordinate vector. Accordingly, it displays the nature of the orbital component $\boldsymbol{J}_{\mathrm{fo}}$ of the angular momentum of the free field

$$\boldsymbol{J}_{\mathrm{fo}} = -\frac{1}{4\pi c}\int_V \mathrm{d}^3r \, \boldsymbol{r} \times \sum_{n=1}^{3} B^n \;\; F^n \quad . \quad (16)$$

By substituting for $\boldsymbol{F}$ from (4) and explicitly taking the gradient this may be expressed as

$$\boldsymbol{J}_{\mathrm{fo}} = \frac{1}{(4\pi c)^2}\int_V \mathrm{d}^3x \int_V \mathrm{d}^3y [\boldsymbol{B}(\boldsymbol{x},t).\frac{\partial \boldsymbol{B}(\boldsymbol{y},t)}{\partial t}] \frac{\boldsymbol{x}\times\boldsymbol{y}}{|\boldsymbol{x}-\boldsymbol{y}|^3} \quad . \quad (17)$$

Equations (15) and (17) or (14) and (16) give the components of the angular momentum of the free fields. The first term of (15) or (14) is a volume integral in $\boldsymbol{x}$ with spin character. The second term of (15) or (14) is a surface integral. Equation (17) is a volume integral with orbital character.

## 3. Plane waves

We apply the above arguments to plane waves. The electric and magnetic fields of a plane wave propagating in the $z$ direction that satisfies Maxwell's equations are given by

$$\boldsymbol{E}(\boldsymbol{x},t) = B[\hat{\boldsymbol{x}}\cos(\omega t - \boldsymbol{k}.\boldsymbol{x} - \alpha) - \hat{\boldsymbol{y}}\cos(\omega t - \boldsymbol{k}.\boldsymbol{x})] \quad (18)$$

$$\boldsymbol{B}(\boldsymbol{x},t) = B[\hat{\boldsymbol{x}}\cos(\omega t - \boldsymbol{k}.\boldsymbol{x}) + \hat{\boldsymbol{y}}\cos(\omega t - \boldsymbol{k}.\boldsymbol{x} - \alpha)] \tag{19}$$

with

$$\frac{\partial \boldsymbol{B}(\boldsymbol{x},t)}{\partial t} = -\omega B[\hat{\boldsymbol{x}}\sin(\omega t - \boldsymbol{k}.\boldsymbol{x}) + \hat{\boldsymbol{y}}\sin(\omega t - \boldsymbol{k}.\boldsymbol{x} - \alpha)] \tag{20}$$

where $\boldsymbol{k} = \hat{\boldsymbol{z}}k$ and $\alpha$ is a constant. For $\alpha = 0$ the magnetic field of the wave is linearly polarized at an angle of $\pi/4$ to the positive $x$ and $y$ axes. For $\alpha = +\pi/2$ the wave is circularly polarized in the $+z$ direction, for $\alpha = -\pi/2$ the wave is circularly polarized in the $-z$ direction. The time average over one cycle of the Poynting vector $\boldsymbol{E}\times\boldsymbol{B}$ is found to be $\hat{\boldsymbol{z}}B^2$, independent of $\alpha$. The time average of the energy density $(\boldsymbol{E}^2 + \boldsymbol{B}^2)/8\pi$ comes to $B^2/4\pi$.

From (4) we calculate the $x$ component of $\boldsymbol{F}$:

$$F^x(\boldsymbol{x},t) = \frac{\omega B}{4\pi c}\int d^3\boldsymbol{x}' \frac{(\sin\omega t\cos\boldsymbol{k}.\boldsymbol{x}' - \cos\omega t\sin\boldsymbol{k}.\boldsymbol{x}')}{|\boldsymbol{x}-\boldsymbol{x}'|} \quad . \tag{21}$$

Using the relations

$$\int d^3\boldsymbol{x}' \frac{\cos\boldsymbol{k}.\boldsymbol{x}'}{|\boldsymbol{x}-\boldsymbol{x}'|} = \frac{4\pi}{k^2}\cos\boldsymbol{k}.\boldsymbol{x} \quad \text{and} \quad \int d^3\boldsymbol{x}' \frac{\sin\boldsymbol{k}.\boldsymbol{x}'}{|\boldsymbol{x}-\boldsymbol{x}'|} = \frac{4\pi}{k^2}\sin\boldsymbol{k}.\boldsymbol{x} \tag{22}$$

we get

$$F^x(\boldsymbol{x},t) = \frac{B}{k}\sin(\omega t - kz) \quad . \tag{23}$$

The $y$ component of $\boldsymbol{F}$ is found in a similar manner to give

$$\boldsymbol{F}(\boldsymbol{x},t) = \frac{B}{k}[\hat{\boldsymbol{x}}\sin(\omega t - \boldsymbol{k}.\boldsymbol{x}) + \hat{\boldsymbol{y}}\sin(\omega t - \boldsymbol{k}.\boldsymbol{x} - \alpha)] \quad . \quad (24)$$

Since $\boldsymbol{k} = \hat{\boldsymbol{z}}k$, $\boldsymbol{F}$ is seen to be a function only of $z$ and $t$. From (19) and (24) the vector cross product of $\boldsymbol{F}$ and $\boldsymbol{B}$ is

$$\boldsymbol{F} \times \boldsymbol{B} = \hat{\boldsymbol{z}}\, B^2 (\sin\alpha) / k \quad . \quad (25)$$

Accordingly, we get for the first term in (14) a contribution to the "spin" angular momentum of

$$\hat{\boldsymbol{z}}\frac{B^2 V}{4\pi\omega}\sin\alpha \quad . \quad (26)$$

where $V$ is the volume of integration. The energy of the wave is $VB^2/4\pi$ so the ratio of spin angular momentum density to energy density is $(\sin\alpha)/\omega$, independent of volume and in agreement with the experiment of Beth [6].

The orbital angular momentum contribution from (16) gives zero because $\boldsymbol{F}$ has a gradient only in the $z$ direction and so the vector cross product of this with $\boldsymbol{r}$ has no component in the $z$ direction.

Finally we examine the surface contribution to the angular momentum. The component of angular momentum in the $z$ direction coming from the second term in (14), the surface term, becomes in Cartesian coordinates

$$J_{\mathrm{fs,\,S}}{}^{z} = \frac{1}{4\pi c}\sum_{r,j,k=1}^{3} \varepsilon^{zjk} \int_S \mathrm{d}^2 x^r (B^r x^j F^k) \quad (27)$$

where $\mathrm{d}^2x^r$ is the element of surface area directed in the $r$ direction. For the present calculation $r$, $j$, $k \neq z$. There are therefore four possible combinations of $r$, $j$ and $k$. First we consider $r = x, j = x,$ $k = y$. This gives a surface integral over the $y$-$z$ planes at $x = +L$ and $x = -L$ as shown shaded in Fig. 1(a)

$$\frac{1}{4\pi c}\int_{-L}^{+L} \mathrm{d}y \int_{-L}^{+L} \mathrm{d}z \, |xB^x F^y|_{-L}^{+L} = \frac{(2L)^2}{4\pi c}\int_{-L}^{+L} \mathrm{d}z \, B^x(z) F^y(z) \quad . \qquad (28)$$

The surface integrals over the two $x$-$z$ planes give a similar result with $-B^yF^x$ replacing $B^xF^y$. The other two surface integrals are zero. Adding the two non-zero terms we get the second term of (14) to be

$$-\frac{(2L)^2}{4\pi c}\int_{-L}^{+L} \mathrm{d}z \, |\boldsymbol{F}(z) \times \boldsymbol{B}(z)|^z \quad . \qquad (29)$$

This is readily seen to be the negative of the contribution given by the first term of (14) so it cancels it to give a total angular momentum of zero.

**6. Discussion**

The angular momentum of a "free" electromagnetic field has been decomposed into three components. The first of them, the first term of (15) or (14), is a volume integral with spin character. The second, the second term of (15) or (14), is a surface integral. The third term (17) is a volume integral with orbital character.

When applied to a plane wave of arbitrary polarization we find that the two terms in (14, 15) are equal and opposite and, when summed, give a total angular momentum of zero. The contribution of (17) is zero also. This finding is independent of the size of the volume of integration and so applies to a plane wave of arbitrary extent. This is clearly consistent with the arguments following equation (1) that assert that a plane wave of arbitrary extent has no angular momentum, but inconsistent with the experimental result [6] that circularly polarized planes wave do carry angular momentum. The resolution of this contradiction becomes apparent when it is realized that such experiments are always performed on beams that are of finite extent because they are constricted by some aperture that is part of the experimental apparatus. The intensity profile of such a beam is as shown in Fig. 1(b). In the central region, the profiles of the intensity and fields approximate those of a plane wave, but at the edges of the beam they drop rapidly to zero in a complicated manner. If the boundaries of the volume over which the integration is performed are placed just outside the beam then the surface integral is zero and the only surviving term is the volume integral. If we had treated the angular momentum arising from the first term of (6) as it stands and had not decomposed it according to (7) we would have got a volume integral whose contribution was dominated by the fields at the edges of the beam. For the case of a plane wave of limited extent, such fields are difficult to calculate and depend on irrelevant details of the properties of the aperture. A much easier calculation is made by using (7) to separate this contribution to the angular momentum into a volume integral and a surface integral. We find that when the effects of boundaries are taken account of the experimental and theoretical properties of plane waves can be reconciled.

Fig. 1(a)

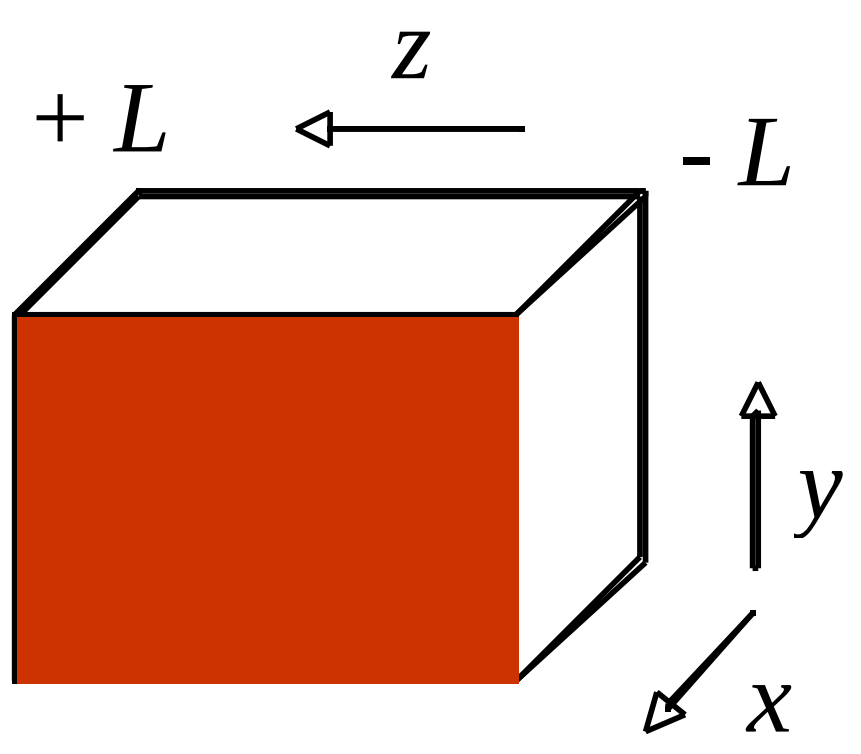

Fig. 1(b)

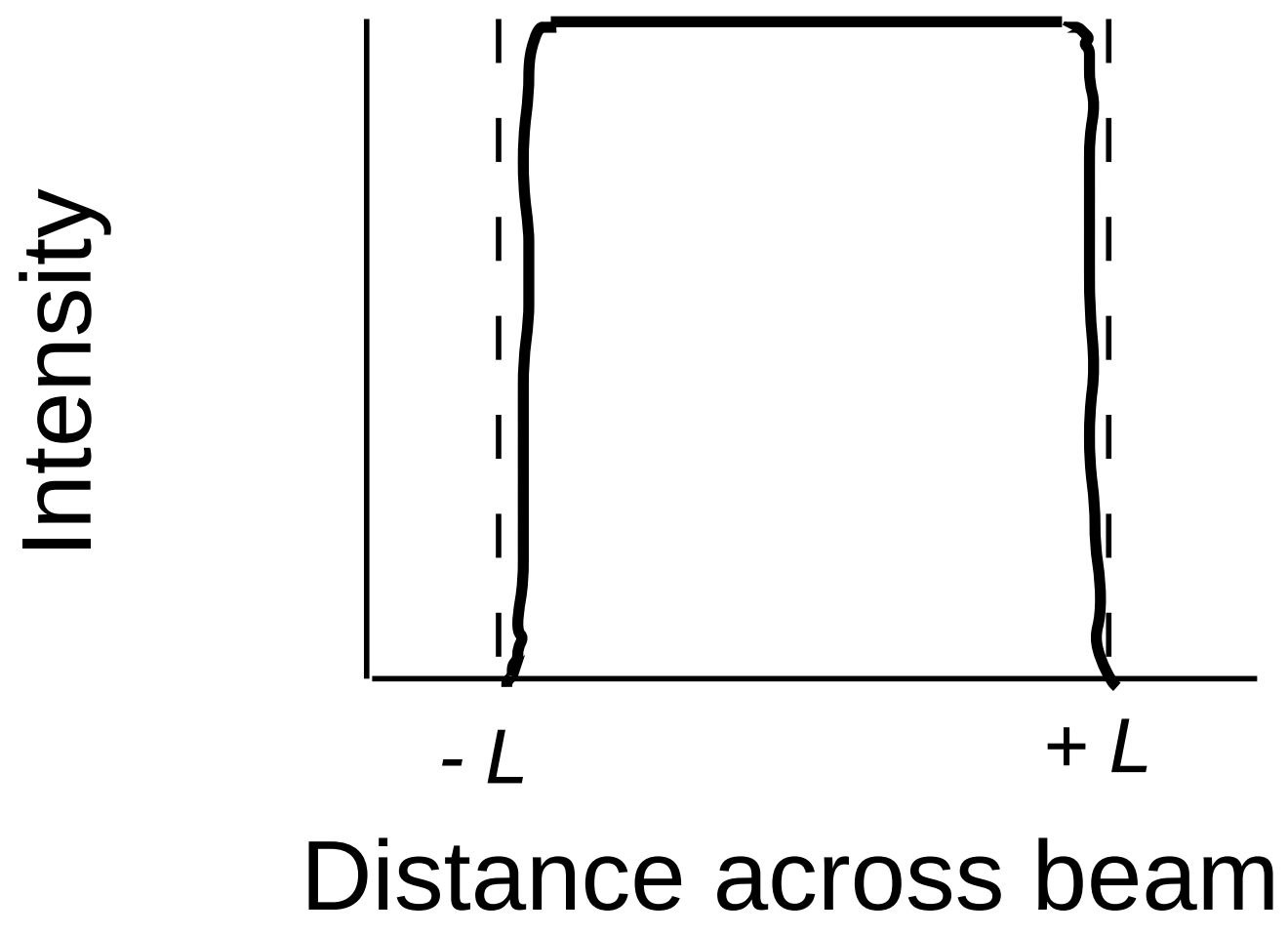

**Figure caption**

Figure 1(a). The volume of integration is a cube of side 2*L* with edges parallel to the coordinate axes.

Figure 1(b). Intensity profile of a plane wave that has passed through an aperture. The central part of the beam closely approximates a plane wave. At the edges of the beam, the fields are cut off relatively sharply and are of a complicated nature.